\documentclass[]{aa}
\usepackage{graphicx}
\usepackage{txfonts}

\begin{document}

\title{Rossby waves in "shallow water" magnetohydrodynamics}

\author{T.V. Zaqarashvili\inst{1,2}, R. Oliver\inst{1}, J.L. Ballester\inst{1} \and B.M. Shergelashvili\inst{2,3}}

\institute{Departament de F\'{\i}sica, Universitat de les Illes
Balears, E-07122 Palma de Mallorca, Spain,
\\ \email{[temury.zaqarashvili;ramon.oliver;dfsjlb0]@uib.es} \and Georgian National
Astrophysical Observatory (Abastumani Astrophysical Observatory),
Kazbegi Ave. 2a, Tbilisi 0160, Georgia, \\ \email{temury@genao.org}
\and Instituut voor Theoretische Fysica, K.U. Leuven,
Celestijnenlaan 200 D, B-3001 Leuven, Belgium,\\
\email{Bidzina.Shergelashvili@fys.kuleuven.be}}

\offprints{T. Zaqarashvili}

\date{Received / Accepted }

\abstract {}{The influence of a toroidal magnetic field on the
dynamics of Rossby waves in a thin layer of ideal conductive fluid
on a rotating sphere is studied in the "shallow water"
magnetohydrodynamic approximation for the first time. }{Dispersion
relations for magnetic Rossby waves are derived analytically in
Cartesian and spherical coordinates.}{It is shown that the magnetic
field causes the splitting of low order (long wavelength) Rossby
waves into two
different modes, here denoted fast and slow {\em magnetic Rossby waves}.
The high
frequency mode (the fast magnetic Rossby mode) corresponds to an
ordinary hydrodynamic Rossby wave slightly modified by the magnetic
field, while the low frequency mode (the slow magnetic Rossby mode) has
new and interesting properties since its frequency is significantly
smaller than that of the same harmonics of pure Rossby and
Alfv{\'e}n waves.} {}


\titlerunning{Magnetic Rossby waves}
\authorrunning{Zaqarashvili et al.}

\maketitle

\section{Introduction}
The large-scale dynamics of planetary atmospheres is mostly
determined by Rossby waves. These waves arise because of the
latitudinal variation of the Coriolis parameter and are widely used
in the geophysical context (Pedlosky \cite{ped}, Gill \cite{gill}).
Rossby waves also can be of importance in solar and stellar
astrophysics, particularly in a thin layer called {\it tachocline}
that is believed to exist below the convection zone of solar-like
stars (Spiegel \& Zahn \cite{spi}, Gough \& McIntyre \cite{gough},
Garaud \cite{garaud}, Cally \cite{cally}, Miesch \cite{mie}). The
thickness of the tachocline is very small compared to the stellar
radius and, therefore, the ordinary shallow water approximation can
be easily applied, but the hydrodynamic (HD) Rossby wave theory
needs to be modified in the presence of a large-scale horizontal
magnetic field. The influence of the horizontal magnetic field on
the large-scale fluid dynamics has been studied in the context of
the Earth's liquid core using the two dimensional $\beta$-plane
approximation in Cartesian coordinates by Hide (\cite{hide}). However,
to study the plasma dynamics over spatial scales comparable to the stellar
radius requires to consider spherical coordinates. Magnetohydrodynamic (MHD)
"shallow water" equations for the solar tachocline have been
recently proposed by Gilman (\cite{gilm}) and the dynamics of
various "shallow water" MHD waves in the solar tachocline have been
studied by Schecter et al. (\cite{schec}) (see also De Sterck
\cite{sterck}). Large-scale Rossby-like waves are absent from their
consideration as the $f$-plane approximation has been used. However,
we should mention the recent work by Leprovost \& Kim
(\cite{leprovost}), which studies the influence of shear, Rossby,
and Alfv\'en waves on the transport properties of MHD turbulence on
a $\beta$-plane in the solar tachocline.

Here we use the MHD "shallow water" equations in order to study the
influence of a toroidal magnetic field on the dynamics of Rossby
waves in a rotating sphere. First we use Cartesian coordinates and
derive the dispersion relation of magnetic Rossby waves in the
$\beta$-plane approximation. Next, we solve the problem in spherical
geometry, thus deriving the propagation properties of magnetic
Rossby waves with a wavelength comparable to the radius of the
sphere.

\section[]{Basic considerations}

Let us consider a thin layer of ideal conductive fluid on a sphere
rotating with angular velocity ${\bf\Omega}_0$. The layer is taken
as an incompressible fluid with a rigid base and a free upper
surface and is permeated by a horizontal uniform magnetic field. The
unperturbed uniform thickness of the layer, $H_0$, is smaller than
the stratification scale height and so the medium density can be
considered uniform. This system differs from the classical shallow
water system only by the presence of the magnetic field. Then, the
"shallow water" MHD equations in an inertial frame can be written as
(Gilman \cite{gilm}, Schecter et al. \cite{schec})
\begin{equation}
{\partial}_t {\bf B}+({\bf V}{\cdot}{\nabla}){\bf B}= ({\bf
B}{\cdot}{\nabla}){\bf V},
\end{equation}
\begin{equation}
{\partial}_t {\bf V}+({\bf V}{\cdot}{\nabla}){\bf V} = {1\over
{4\pi\rho}}({\bf B}{\cdot}{\nabla}){\bf B}-g{\nabla H},
\end{equation}
\begin{equation}
{\partial}_t {H}+ {\nabla}{\cdot}(H{\bf V})=0,
\end{equation}
where $\bf V$ and $\bf B$ are the horizontal velocity and magnetic
field, $H$ is the thickness of the layer, $\rho$ is the fluid
density, $\nabla$ is the horizontal gradient and $g$ is the
gravitational acceleration. The divergence-free condition for the
magnetic field, which arises from the requirement that $\bf B$ is
parallel to the upper free surface, takes the form (Gilman
\cite{gilm})
\begin{equation}
{\nabla}{\cdot}(H{\bf B})=0.
\end{equation}

A general feature of the large-scale dynamics in such a system is
that it does not significantly depend on the chosen geometry (flat,
spherical or cylindrical) for equatorially trapped waves
(Longuet-Higgins \cite{long}, Pedlosky \cite{ped}). However, the
consideration of spherical geometry is desirable for waves with
wavelength comparable to the size of the sphere. Therefore, we first
study the problem in the simpler Cartesian coordinates and then turn to
the more complicate spherical geometry.

\section{Cartesian coordinates}

We consider a local Cartesian coordinate frame $(x,y,z)$ in which
the $x$ axis is directed towards the rotation, the $y$ axis is
directed towards the north pole of the sphere and the $z$ axis is
directed vertically.

Let us next consider that the unperturbed magnetic field,
$(B_x,0,0)$, is directed along the $x$ axis. Then, after linearizing
Eqs.~(1)--(3) their components are written in the rotating
frame as

\begin{equation}
{{\partial u_x}\over {\partial t}} - fu_y={{B_x}\over
{{4\pi\rho}}}{{\partial b_x}\over {\partial x}}-g{{\partial
h}\over {\partial x}},
\end{equation}
\begin{equation}
{{\partial u_y}\over {\partial t}} + fu_x={{B_x}\over
{{4\pi\rho}}}{{\partial b_y}\over {\partial x}}-g{{\partial
h}\over {\partial y}},
\end{equation}
\begin{equation}
{{\partial b_x}\over {\partial t}}=B_x{{\partial u_x}\over {\partial
x}},\,\,\, {{\partial b_y}\over {\partial t}}=B_x{{\partial
u_y}\over {\partial x}},
\end{equation}
\begin{equation}
{{\partial h}\over {\partial t}} + H_0\left ({{\partial u_x}\over
{\partial x}}+ {{\partial u_y}\over {\partial y}}\right )=0,
\end{equation}
where $u_x$, $u_y$, $b_x$ and $b_y$ are the velocity and magnetic
field perturbations, $h=H-H_0$ is the perturbation of the layer
thickness and $f=2{\Omega_0}\sin{\Theta}$ is the Coriolis parameter
(with $\Theta$ the latitude). For zero magnetic field this system
transforms into the HD shallow water equations (Pedlosky
\cite{ped}).

Differentiation with respect to time of Eqs.~(5)--(6) and using
Eqs.~(7)--(8) gives
\begin{equation}
{{\partial^2 u_x}\over {\partial t^2}} - f{{\partial u_y}\over
{\partial t}}=v^2_A{{\partial^2 u_x}\over {\partial x^2}}+C^2_0
\left ({{\partial^2 u_x}\over {\partial x^2}}+ {{\partial^2
u_y}\over {{\partial x}{\partial y}}}\right ),
\end{equation}
\begin{equation}
{{\partial^2 u_y}\over {\partial t^2}} + f{{\partial u_x}\over
{\partial t}}=v^2_A{{\partial^2 u_y}\over {\partial x^2}}
+C^2_0\left ({{\partial^2 u_x}\over {{\partial x}{\partial y}}} +
{{\partial^2 u_y}\over {\partial y^2}}\right ),
\end{equation}
where $v_A=B_x/\sqrt{4\pi \rho}$ and $C_0=\sqrt{gH_0}$ are the
Alfv{\'e}n and surface gravity speeds, respectively.

We now perform a Fourier analysis of the form $\exp(-i\omega t +
ik_x x)$ and after some algebra obtain
\begin{eqnarray}
{{\partial^2 u_y}\over {\partial y^2}} + \left [{{\omega^2}\over
{C^2_0}}-k^2_x -{{k^2_xv^2_A}\over {C^2_0}} - {{\omega^2 f^2}\over
{C^2_0(\omega^2-k^2_xv^2_A)}}- {{k_x\omega}\over
{(\omega^2-k^2_xv^2_A)}}{{\partial f}\over {\partial y}}\right
]u_y\nonumber \\ =0.
\end{eqnarray} When $v_A=0$ this equation governs the linear
dynamics of various kinds of waves (namely Poincar{\'e}, Kelvin and
Rossby waves) in the HD shallow water approximation (Pedlosky
\cite{ped}), but the inclusion of the magnetic field leads to the
modification of the wave modes.

At a given latitude, $\Theta_0$, one can perform a Taylor expansion
of the Coriolis parameter and retain the lowest order latitudinal
variation of $f$, which leads to (Pedlosky \cite{ped}, Gill
\cite{gill})
\begin{equation}
f=f_0 + \beta y,
\end{equation}
where the parameter
\begin{equation}
\beta={{2\Omega_0}\over R_0}\cos{\Theta_0}
\end{equation}
(with $R_0$ the radius of the sphere) plays a major role in the so called
$\beta$-plane approximation.
Away from the equator $\beta y \ll f_0$ and therefore from Eq.~(11) we readily get
\begin{eqnarray}
{{\partial^2 u_y}\over {\partial y^2}} + \left [{{\omega^2}\over
{C^2_0}}-k^2_x -{{k^2_xv^2_A}\over {C^2_0}}- {{\omega^2 f^2_0}\over
{C^2_0(\omega^2-k^2_xv^2_A)}} -{{k_x\omega \beta}\over
{(\omega^2-k^2_xv^2_A)}}\right ]u_y \nonumber \\ =0. \end{eqnarray}
Thus, we can now perform a Fourier analysis of the form $\exp(ik_y
y)$, which gives the dispersion relation
\begin{eqnarray}
\omega^4 - [2k^2_xv^2_A +f^2_0+C^2_0(k^2_x +k^2_y)]\omega^2 -
C^2_0k_x\beta\omega + \nonumber \\ k^2_xv^2_A[k^2_xv^2_A
+C^2_0(k^2_x +k^2_y)]
 =0.
\end{eqnarray} This dispersion relation contains high and low
frequency branches, which respectively correspond to magneto-gravity
waves and to Alfv{\'e}n and Rossby waves. Note that for $\beta=0$
this dispersion relation transforms into that of $f$-plane MHD
"shallow water" waves (Schecter et al. \cite{schec}).

We next concentrate in the case of small Alfv{\'e}n speed, i.e. $v_A
\ll C_0$, such as corresponds to the interiors of solar-like stars. Then,
the high frequency branch of Eq.~(15) contains Poincar{\'e}
waves, whose dispersion relation is  (Pedlosky \cite{ped})
\begin{equation}
\omega^2 = f^2_0+C^2_0(k^2_x +k^2_y),
\end{equation}
while the low frequency branch yields the dispersion relation
\begin{equation}
\omega^2 + {{k_x\beta}\over {k^2_x +k^2_y}}\omega - k^2_xv^2_A=0.
\end{equation}
Note that the dispersion relation (17) was first obtained by Hide
(\cite{hide}) in the two dimensional case (see also Acheson \& Hide
\cite{ach}). This formula reveals some interesting properties. For
short wavelengths, i.e. large $k_x$, the last term in Eq.~(17)
dominates over the second one, which leads to the solution
\begin{equation}
\omega = \pm k_xv_A.
\end{equation}
This is the dispersion relation of pure Alfv\'en waves unaffected by
rotation and propagating eastward and westward in the toroidal
direction.

Nevertheless, for large-scale motions pure Alfv{\'e}n waves no longer
exist and instead we have Rossby waves modified by the magnetic field.
For large wavelengths, i.e. small $k_x$, Eq.~(17) has two different solutions. For
the high frequency solution one can easily recover the dispersion
relation of HD Rossby waves,
\begin{equation}
\omega \approx -{{k_x\beta}\over {k^2_x +k^2_y}}.
\end{equation}
For the low frequency solution we have the dispersion relation
\begin{equation}
\omega \approx {{k_xv^2_A(k^2_x +k^2_y)}\over {\beta}}.
\end{equation}
Hence, the horizontal magnetic field causes the splitting of
ordinary large-scale Rossby waves into two modes propagating in
opposite directions. The high frequency mode has the properties of
HD Rossby waves and can be called {\it fast magnetic Rossby mode}.
But, additionally, a lower frequency mode arises whose frequency is
significantly smaller than that of pure Alfv{\'e}n and Rossby waves
at the same spatial scale. Due to its small frequency it can be
called {\it slow magnetic Rossby mode} ("hydromagnetic-planetary
waves" in Acheson \& Hide \cite{ach}). The phase speed of the mode
in the $x$ direction depends on both the Alfv{\'e}n speed and the
$\beta$ parameter,
\begin{equation}
v_{ph}={{\omega}\over k_x} ={{v^2_A(k^2_x +k^2_y)}\over {\beta}}.
\end{equation}
The phase speed is different from Alfv{\'e}n and Rossby phase
speeds, which again indicates the different nature of this wave
mode.

\begin{figure}
\centering
\includegraphics[width=0.8\linewidth]{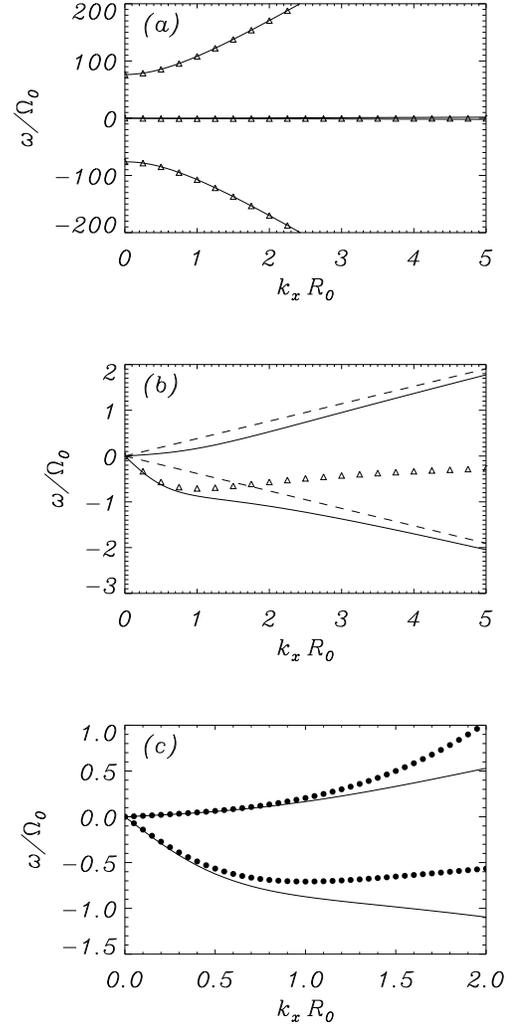}
\caption{(a) Dispersion diagram of "shallow water" waves in the
presence of a horizontal magnetic field. The two extreme upper and
lower solutions correspond to Poincar{\'e} waves, which are almost
unaffected by the magnetic field, whereas the two low frequency
modes are magnetic Rossby waves. (b) Low frequency branch of the
dispersion diagram. (c) Low frequency, large wavelength part of the
dispersion diagram. In all panels solid lines correspond to the
solutions to Eq.~(15), triangles to the HD Poincar\'e and Rossby
waves (i.e. in the absence of magnetic field), dashed lines to pure
Alfv{\'e}n waves (whose dispersion relation is given by Eq.~(18))
and circles to the approximate analytical formulas (19)--(20). The
parameters used to obtain the dispersion diagram are $k_y R_0=1$,
$v_A/C_0=0.005$ and $\Theta=45^\circ$. \label{fig1}}
\end{figure}

Numerical dispersion diagrams for the general dispersion relation
(15),  i.e. without assuming $v_A\ll C_0$,
are presented in Fig.~1. The upper panel displays all wave
solutions and shows that Poincar{\'e} waves are almost not affected
by the magnetic field. The middle panel is a detailed view of the
low frequency branch of the dispersion diagram. It is clearly seen
that for small scales, i.e. for large $k_x$, magnetic Rossby waves
tend to the Alfv{\'e}n wave solutions (dashed lines), whereas for
small $k_x$, i.e. for large spatial scales, the two modes behave
differently: the solution with higher negative frequency corresponds
to HD Rossby waves (triangles) and the low frequency solution, which
differs from pure Rossby and Alfv{\'e}n wave dispersion curves, is a
new wave mode. Finally, the bottom panel shows the perfect fit
between the solutions to Eq.~(15) and the approximate
dispersion relation for magnetic Rossby waves (circles),
Eqs.~(19)--(20), in the limit of small $k_x$.

The present consideration in Cartesian coordinates gives the basic
properties of magnetic Rossby waves. Nevertheless, to study the
dynamics of Rossby waves with spatial scales comparable to the
radius of the sphere it is desirable to use spherical coordinates.
Therefore, in the next section we study the same problem in
spherical coordinates ($r,\theta, \phi$).

\section{Spherical coordinates}

Let us consider an unperturbed toroidal magnetic field $B_{\phi}$. Then,
the linearized form of Eqs.~(1)--(3) can be rewritten in the rotating frame as
\begin{eqnarray}
{{\partial u_{\theta}}\over {\partial t}} - 2\Omega_0 \cos \theta
u_{\phi} +{{g}\over R_0}{{\partial h}\over {\partial \theta}}-
{{B_{\phi}}\over {{4\pi\rho R_0 \sin \theta}}}{{\partial
b_{\theta}}\over {\partial \phi}}+ \nonumber \\
2 {{B_{\phi}}\over {{4\pi\rho R_0 }}}{{\cos \theta}\over {\sin
\theta}}b_{\phi} =0, \end{eqnarray}
\begin{eqnarray}
{{\partial u_{\phi}}\over {\partial t}} + 2\Omega_0 \cos \theta
u_{\theta} +{{g}\over {R_0 \sin \theta}}{{\partial h}\over {\partial
\phi}}- {{b_{\theta}}\over {{4\pi\rho R_0}}}{{\partial
B_{\phi}}\over {\partial \theta}}- \nonumber
\\ {{B_{\phi}}\over {{4\pi\rho R_0
\sin \theta}}}{{\partial b_{\phi}}\over {\partial \phi}}-
{{B_{\phi}}\over {{4\pi\rho R_0 }}}{{\cos \theta}\over {\sin
\theta}}b_{\theta}=0, \end{eqnarray}
\begin{equation}
{{\partial h}\over {\partial t}} + {{H_0}\over {R_0 \sin
\theta}}{{\partial }\over {\partial \theta}}\left (\sin \theta
u_{\theta} \right ) + {{H_0}\over {R_0 \sin \theta}}{{\partial
u_{\phi}}\over {\partial \phi}}=0,
\end{equation}
\begin{equation}
{{\partial b_{\theta}}\over {\partial t}}- {{B_{\phi}}\over {{R_0
\sin \theta}}}{{\partial u_{\theta}}\over {\partial \phi}}=0,
\end{equation}
\begin{equation}
{{\partial b_{\phi}}\over {\partial t}}+ {{1}\over
{{R_0}}}{{\partial }\over {\partial \theta}}\left
(u_{\theta}B_{\phi} \right )=0,
\end{equation}
where $u_{\theta}$, $u_{\phi}$, $b_{\theta}$ and $b_{\phi}$ are the
velocity and magnetic field perturbations, while $h=H-H_0$ is the
perturbation of the layer thickness. For zero magnetic field this
system transforms into the HD shallow water equations
(Longuet-Higgins \cite{long}).

We assume the unperturbed magnetic field to be $B_{\phi}=B_0
\sin{\theta}$, which means that it has a maximal value at the
equator and tends to zero at the poles. We take a sinusoidal
dependence of the magnetic field on $\theta$ for two main reasons:
first, the sinusoidal profile simplifies the calculation in the
spherical symmetry and second, the toroidal magnetic field seems to
be located mainly in low latitudes due to the eruption of magnetic
flux at these latitudes. Let us next introduce the new variables
${\hat u}_{\theta}= \sin{\theta}u_{\theta}$, ${\hat u}_{\phi}=
\sin{\theta}u_{\phi}$, ${\hat b}_{\theta}= \sin{\theta}b_{\theta}$,
${\hat b}_{\phi}= \sin{\theta}b_{\phi}$. Then, Eqs.~(22)--(26) take
the form:
\begin{eqnarray}
{{\partial {\hat u}_{\theta}}\over {\partial t}} - 2\Omega_0 \cos
\theta {\hat u}_{\phi} +{{g}\over R_0}\sin \theta{{\partial h}\over
{\partial \theta}}- {{B_{0}}\over {{4\pi\rho R_0}}}{{\partial {\hat
b}_{\theta}}\over {\partial \phi}}+ \nonumber \\ 2{{B_{0}}\over
{{4\pi\rho R_0 }}}{\cos \theta}{\hat b}_{\phi} =0, \end{eqnarray}
\begin{equation}
{{\partial {\hat u}_{\phi}}\over {\partial t}} + 2\Omega_0 \cos
\theta {\hat u}_{\theta} +{{g}\over {R_0}}{{\partial h}\over
{\partial \phi}}-{{B_{0}}\over {{4\pi\rho R_0}}}{{\partial {\hat
b}_{\phi}}\over {\partial \phi}}- {{2B_{0}}\over {{4\pi\rho R_0
}}}{\cos \theta}{\hat b}_{\theta}=0,
\end{equation}
\begin{equation}
\sin^2 \theta{{\partial h}\over {\partial t}} + {{H_0}\over {R_0
}}\sin \theta{{\partial {\hat u}_{\theta}}\over {\partial \theta}} +
{{H_0}\over {R_0}}{{\partial {\hat u}_{\phi}}\over {\partial
\phi}}=0,
\end{equation}
\begin{equation}
{{\partial {\hat b}_{\theta}}\over {\partial t}}- {{B_{0}}\over
{{R_0}}}{{\partial {\hat u}_{\theta}}\over {\partial \phi}}=0,
\end{equation}
\begin{equation}
{{\partial {\hat b}_{\phi}}\over {\partial t}}+ {{B_0}\over
{{R_0}}}\sin \theta{{\partial {\hat u}_{\theta}}\over {\partial
\theta}}=0.
\end{equation}

We now perform a Fourier analysis of the form $\exp(-i\omega t
+ is\phi)$ and define
\begin{eqnarray}
{{\omega}\over {2\Omega_0}}=\lambda,\,{{4\Omega^2_0R^2_0}\over {g
H_0}}=\epsilon,\,{{v^2_A}\over {4\Omega^2_0R^2_0}}=\alpha^2,\ {{g
h}\over {2\Omega_0 R_0}}=\eta,\, \nonumber\\ \cos \theta=\mu,\,\,
-\sin \theta{{\partial }\over {\partial \theta}}=(1-\mu^2){{\partial
}\over {\partial \mu}}=D.
\end{eqnarray}
After some algebra we get
\begin{equation}
-\lambda^2{\tilde u}_{\theta}-\mu\lambda{\hat u}_{\phi}-\lambda
D\eta +s^2\alpha^2{\tilde u}_{\theta} + 2\alpha^2\mu D{\tilde
u}_{\theta}=0,
\end{equation}
\begin{equation}
-\lambda^2{\hat u}_{\phi}-\mu\lambda{\tilde u}_{\theta}+\lambda
s\eta -2s\alpha^2\mu {\tilde u}_{\theta} - \alpha^2s D{\tilde
u}_{\theta}=0,
\end{equation}
\begin{equation}
\epsilon \lambda(1-\mu^2)\eta - D{\tilde u}_{\theta}-s{\hat
u}_{\phi}=0,
\end{equation}
where ${\tilde u}_{\theta}=i{\hat u}_{\theta}$.

Substitution of ${\hat u}_{\phi}$ from Eq.~(34) into Eqs.~(33) and
(35) leads to
\begin{eqnarray}
-\lambda^2{\tilde u}_{\theta}+\mu^2{\hat u}_{\phi}-(\lambda
D+s\mu)\eta +s^2\alpha^2{\tilde u}_{\theta} + 2\alpha^2\mu D{\tilde
u}_{\theta} + \nonumber \\
\mu s {{\alpha^2}\over \lambda}(D+2\mu){\tilde u}_{\theta}=0,
\end{eqnarray}
\begin{equation}
\epsilon \lambda(1-\mu^2)\eta - D{\tilde u}_{\theta}+{s\over
\lambda}(\mu{\tilde u}_{\theta}-s\eta) + {{\alpha^2}\over
{\lambda^2}}s^2(D +2\mu){\tilde u}_{\theta}=0.
\end{equation}

\noindent After obtaining $\eta$ from Eq. (37) and substituting it into
Eq.~(36) we get a single equation for ${\tilde u}_{\theta}$,
\begin{eqnarray}
(\lambda D +s \mu)\left \{{1\over {s^2 -\epsilon
\lambda^2(1-\mu^2)}}\left [\lambda D -s \mu- {{\alpha^2}\over
{\lambda^2}}s^2\lambda(D +2\mu)\right ]\right \}{\tilde u}_{\theta}
\nonumber \\
-(\lambda^2 -\mu^2){\tilde u}_{\theta}+s^2\alpha^2{\tilde
u}_{\theta} +2\alpha^2\mu D{\tilde u}_{\theta} +\mu s {\alpha^2\over
\lambda}(D+2\mu){\tilde u}_{\theta}=0.
\end{eqnarray}

In the approximation for slowly rotating stars (such as in the solar
case),
\begin{equation}
{{\epsilon}\over {s^2}}={{4\Omega^2_0R^2_0}\over {g H_0 s^2}}\ll 1
\end{equation}
and Eq.~(38) takes the form
\begin{eqnarray}
[(\lambda D +s\mu)(\lambda D -s\mu) - {{\alpha^2}\over
{\lambda}}s^2(\lambda D +s\mu)(D +2\mu) - \nonumber \\ s^2(\lambda^2
-\mu^2)+ s^4\alpha^2 + 2\alpha^2 s^2 \mu D +\mu s^3 {{\alpha^2}\over
\lambda}(D +2\mu)]{\tilde u}_{\theta}=0.
\end{eqnarray}

\noindent The approximation (39) implies that magneto gravity waves
(i.e. magnetic
Poincar{\'e} and Kelvin waves) are neglected from our consideration, thus
retaining only magnetic Rossby waves.

Now, Eq.~(40) can be rewritten as
\begin{equation}
\left [{{\partial }\over {\partial \mu}}(1-\mu^2){{\partial }\over
{\partial \mu}} - {s^2\over {1-\mu^2}}+ n(n+1) \right ]{\tilde
u}_{\theta}=0,
\end{equation}
if
\begin{equation}
n(n+1)= -{{s\lambda + 2s^2\alpha^2}\over {\lambda^2-\alpha^2s^2}}.
\end{equation}

\noindent Equation~(41) is the associated Legendre differential equation
(Abramowitz \& Stegun \cite{abr}), whose typical solutions are the
associated Legendre polynomials,
\begin{equation}
{\tilde u}_{\theta}=P^{s}_n(\cos \theta)
\end{equation}
if $n$ is an integer (when $n$ is not integer the solutions are the
associated Legendre functions).

Equation~(42) defines the dispersion relation for spherical magnetic
Rossby waves,
\begin{equation}
n(n+1)\left ({{\lambda}\over s}\right )^2 + {{\lambda}\over s} +
\alpha^2[2-n(n+1)]=0,
\end{equation}
where $n$ plays the role of the poloidal wavenumber.

In the non-magnetic case, i.e. for $\alpha=0$, the dispersion relation reduces
to the HD Rossby mode solution (Longuet-Higgins \cite{long}). But
the magnetic field causes the splitting of the ordinary HD mode into
fast and slow magnetic Rossby modes as in the rectangular case (it is worth
recalling that Rossby-type waves are sometimes called planetary waves,
which is probably a more appropriate name for the spherical geometry).

Before studying the dispersion diagram, let us first consider the
properties of wave modes for particular values of $n$. For purely
toroidal propagation, i.e. when $n=0$, we have only one mode with
the dispersion relation
\begin{equation}
{{\omega}\over s}= - {{v^2_A}\over {\Omega_0 R^2_0}}.
\end{equation}
In the weak magnetic field limit this mode has significantly lower
frequency than pure Rossby and Alfv{\'e}n waves and can be
associated to the slow magnetic Rossby mode. Note, that $n=0$ mode
is absent in non-magnetic spherical case (Longuet-Higgins
\cite{long}).

For $n=1$ we have only the fast magnetic Rossby mode, which in this
case is identical to the HD Rossby mode, with the dispersion
relation (Longuet-Higgins \cite{long})
\begin{equation}
{{\omega}\over s} =  - \Omega_0.
\end{equation}

\begin{figure}
\centering
\includegraphics[width=0.8\linewidth]{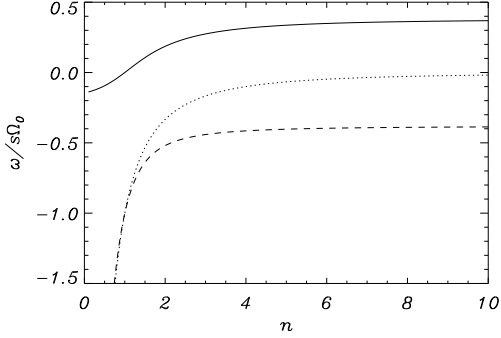}
\caption{Numerical dispersion diagram of spherical "shallow water"
waves in the presence of a toroidal magnetic field. The dependence of
the wave frequency on the poloidal wavenumber $n$ is plotted here
for $\alpha^2=0.036$. The continuous and dashed lines are the
solutions for the slow and fast magnetic Rossby modes, respectively. The
dotted line is the HD Rossby mode, which is obtained from Eq. (44)
with $\alpha^2=0$.
 \label{fig2}}
\end{figure}

For $n>1$ we have both fast and slow magnetic Rossby modes. In the
weak magnetic field limit, i.e. for $v_A \ll 2\Omega_0 R_0$, the
dispersion relation for the fast magnetic Rossby mode is
\begin{equation}
{{\omega}\over s} \approx - {{2\Omega_0}\over {n(n+1)}}
\end{equation}
(which is similar to HD Rossby mode, Longuet-Higgins \cite{long}) and the
dispersion relation for the slow magnetic Rossby mode is
\begin{equation}
{{\omega}\over s} \approx - {{v^2_A}\over {\Omega_0
R^2_0}}[2-n(n+1)].
\end{equation}

\noindent
Equations~(47)--(48) show that the toroidal magnetic field causes the
splitting of ordinary HD Rossby waves into two different modes
propagating in opposite directions for $n>1$. Such splitting is also
present for all non-integer values of $n$ smaller than 1. Thus, the general
behaviour of the wave modes in spherical geometry is similar to
that of the Cartesian case. There are some differences, however.
First, the dispersion relation of magnetic Rossby waves in a
Cartesian frame depends on the latitude; second, magnetic Rossby
waves are dispersive with respect to the toroidal wavenumber, $k_x$,
in Cartesian geometry (see Eqs.~(19)--(20)), while they are not
dispersive with respect to the toroidal wavenumber, $s$, in spherical
geometry.

\begin{figure}
\centering
\includegraphics[width=0.8\linewidth]{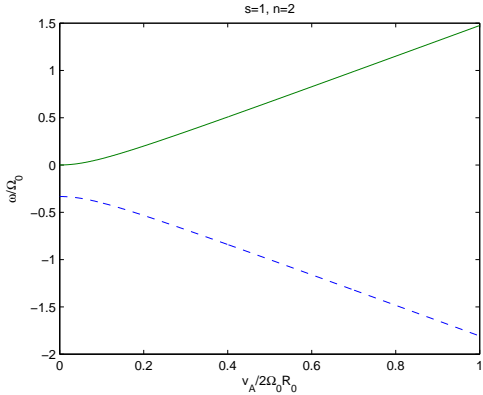}
\caption{Frequencies of the $s=1, n=2$ harmonics of fast (dashed) and
slow (continuous) magnetic Rossby modes vs. the ratio of the
Alfv{\'e}n speed to the rotation rate, $\alpha=v_A/2\Omega_0 R_0$.
When $\alpha$ tends to zero, then the slow magnetic
Rossby mode has a very low frequency, while the fast mode frequency
tends to $\sim 0.3\Omega_0$.
 \label{fig3}}
\end{figure}

Numerical solutions to the general dispersion relation (44) are
presented in Fig. 2. The dependence of the wave frequency on the
poloidal wavenumber $n$ is plotted here for $\alpha^2=0.036$. The
continuous and dashed lines are the solutions for slow and fast
magnetic Rossby modes, respectively. The dotted line is the HD
Rossby mode, which is obtained from Eq. (44) with $\alpha^2=0$. It
is clearly seen that for small scales, i.e. for large $n$, fast and
slow magnetic Rossby waves behave similarly and tend to the
Alfv{\'e}n-like wave solutions, whereas for $n<3$, i.e. for large
spatial scales, the two modes behave differently: the solution with
higher negative frequency corresponds to HD Rossby waves (dashed
line) and the low frequency (continuous line) solution, which
differs from pure Rossby wave dispersion curves, is a new wave mode
(see similar consideration by Hide \cite{hide}, Acheson \& Hide
\cite{ach}). It must be mentioned that the value of the toroidal
wavenumber, $s$, does not influence the general behaviour of the
modes and that increasing $s$ leads only to a linear increase of the
frequency, $\omega$.

The frequency of particular harmonics depends on the strength of the
toroidal magnetic field. The frequency of the $s=1, n=2$ harmonics of
fast (dashed) and slow (continuous) magnetic Rossby modes vs. the
ratio of the Alfv{\'e}n speed to the rotation rate,
$\alpha=v_A/2\Omega_0 R_0$, are plotted in Fig. 3. The fast and slow
modes have lower frequency than the angular velocity, $\Omega_0$, for
$\alpha<0.5$ and $\alpha<0.8$ respectively. When $\alpha$ goes to
zero, the fast mode frequency tends to
$0.3\Omega_0$, while the slow mode frequency tends to zero.
Therefore, the slow magnetic Rossby mode may have very low frequency
depending on the ratio of the Alfv{\'e}n speed to the angular
velocity.

\section{Discussion}

The recently developed "shallow water" magnetohydrodynamic approximation
(Gilman \cite{gilm}) has stimulated further study of MHD wave modes in
this system. While ordinary "shallow water" modes (Poincar{\'e},
Kelvin, Rossby) have been intensively studied in the geophysical
context (Pedlosky \cite{ped}, Gill \cite{gill}), the inclusion of
a horizontal magnetic field enriches the wave spectrum. Magneto-gravity
and Alfv\'en modes in the "shallow water" MHD system have been
recently studied by Schecter et al. (\cite{schec}). However,
large-scale modes (those with stellar spatial dimensions; for
example, Rossby waves) were absent from their consideration due to
the use of the $f$-plane approximation. On the contrary, here we emphasize the
large-scale behavior of wave modes in the "shallow water" system.
Considering the $\beta$-plane approximation and, especially, spherical
coordinates, enables us to study the wave dynamics on spatial scales
corresponding to stellar dimensions. Particular attention is paid to
the slow magnetic Rossby mode, expressed by Eqs. (20),
(45), (48). The presence of this mode in rotating magnetised fluids
was first pointed out by Hide ({\cite{hide}}) using the two dimensional Cartesian
$\beta$-plane approximation in the context of the Earth's liquid core. Here 
we derive analytical dispersion relations of this mode in both Cartesian and spherical "shallow
water" MHD systems. In the low Alfv\'en speed
limit (compared to the surface gravity speed), this mode has a smaller
frequency than that of pure Alfv\'en and Rossby modes and
consequently may have new interesting consequences in large-scale stellar
dynamics.

However, this consideration needs some modifications when applied to
concrete astrophysical situations; for example, to the solar
tachocline. It is believed that the tachocline is divided in two
parts: the inner "radiative" layer with a strongly stable
stratification and the outer "overshoot" layer with a weakly stable
stratification (Gilman \cite{gilm}). The subadiabatic stratification
provides negative buoyancy in both layers, which leads to the so-called
"reduced gravity", $g_r$ (Gilman \cite{gilm}; Schecter et al.
\cite{schec}). Therefore, the developed theory of fast and slow
magnetic Rossby waves should be modified for tachoclines of
solar-like stars using the reduced gravity instead of the ordinary
one. Then the results can be quite different for the radiative and
overshoot layers due to the significant difference between the
reduced gravity there. Schecter et al. (\cite{schec}) estimated the
reduced gravity in the radiative layer as 500--1.5 $\cdot$ 10$^4$
cm$\cdot$s$^{-2}$ and in the overshoot layer as 0.05--5
cm$\cdot$s$^{-2}$. Then the surface gravity speed $C_0$ is higher
than the Alfv\'en speed in the radiative layer, but not in the
overshoot one, where both speeds may have similar values
(Schecter et al. \cite{schec}). Therefore, the results obtained here 
can be easily applied to the radiative part of
tachocline, but not to the overshoot one. The exception is the 
dispersion relation (15), which can be applied to both parts of the tachocline.

It must be also mentioned that differential rotation, typical of
the tachocline dynamics, is absent from our consideration. The goal
of this paper is to study the influence of the magnetic field on the
large-scale dynamics of a "shallow water" system in general. However,
differential rotation should be taken into account in more
realistic models of wave dynamics in the tachocline. 

\section{Conclusions}

The influence of a toroidal magnetic field on the dynamics of Rossby
waves in a thin layer of ideal conductive fluid on a rotating sphere
is studied in the shallow water MHD approximation for both, Cartesian and 
spherical geometries. It is shown that
in both cases the magnetic field causes the splitting of low order
(long wavelength)
ordinary Rossby wave harmonics into two modes (here called magnetic
Rossby modes). The high frequency
mode (the fast magnetic Rossby mode) corresponds to ordinary HD Rossby
waves slightly modified by the magnetic field, while the low
frequency solution leads to a new mode (the slow magnetic Rossby mode)
with interesting properties. Low order (with respect to the
poloidal wavenumber) harmonics of the slow magnetic Rossby mode have
lower frequency than the pure Rossby and Alfv{\'e}n wave frequencies
of the corresponding harmonics.

\begin{acknowledgements}

This work has been supported by MEC grant AYA2006-07637 and the
Georgian National Science Foundation grant GNSF/ST06/4-098. The work
of B.M.S. has been supported by a postdoctoral fellowship in the
Institute for Theoretical Physics, K.U. Leuven - PDM/06/116. We thank the
referee for some helpful comments.

\end{acknowledgements}

{}

\end{document}